 \documentclass[aps,prl,twocolumn,superscriptaddress,showpacs]{revtex4}

\usepackage{graphicx}
\usepackage{dcolumn}
\usepackage{bm}

\begin{document}

\title{Diameter-dependent conductance oscillations in
       carbon nanotubes upon torsion}

\author{K. S. Nagapriya}
\affiliation{Department of Materials and Interfaces,
             Weizmann Institute of Science,
             Rehovot 76100, Israel}

\author{Savas Berber}
\affiliation{Physics and Astronomy Department,
             Michigan State University,
             East Lansing, Michigan 48824-2320 }

\author{Tzahi Cohen-Karni}
\altaffiliation[Current address: ]
             {Harvard University,
             12 Oxford Street,
             Cambridge, MA 02138}
\affiliation{Department of Materials and Interfaces,
             Weizmann Institute of Science,
             Rehovot 76100, Israel}

\author{Lior~Segev}
\affiliation{Department of Materials and Interfaces,
             Weizmann Institute of Science,
             Rehovot 76100, Israel}

\author{Onit Srur-Lavi}
\affiliation{Department of Materials and Interfaces,
             Weizmann Institute of Science,
             Rehovot 76100, Israel}

\author{David Tom\'anek}
\email[e-mail: ]{tomanek@pa.msu.edu}
\affiliation{Physics and Astronomy Department,
             Michigan State University,
             East Lansing, Michigan 48824-2320 }

\author{Ernesto Joselevich}
\email[e-mail: ]{ernesto.joselevich@weizmann.ac.il}
\affiliation{Department of Materials and Interfaces,
             Weizmann Institute of Science,
             Rehovot 76100, Israel}


\begin{abstract}
We combine electromechanical measurements with {\em ab initio}
density functional calculations to settle the controversy about
the origin of torsion-induced conductance oscillations in
multi-wall carbon nanotubes. According to our observations, the
oscillation period is inversely proportional to the squared
diameter of the nanotube, as expected for a single-wall nanotube
of the same diameter. This is supported by our theoretical finding
that differential torsion effectively decouples the walls of a
multi-wall nanotube near the Fermi level and moves the Fermi
momentum across quantization lines. We exclude the alternative
explanation linked to registry changes between the walls, since it
would cause a different diameter dependence of the oscillation
period.
\end{abstract}

\date{\today}




\maketitle



Carbon nanotubes~\cite{TAP111} (CNTs) are mechanically robust and
electrically conducting, and thus seem well suited for use in
nanoelectromechanical systems
(NEMS)~\cite{stiffening,Superfine1,Fennimore, Superfine2,Meyer}.
The electronic response of CNTs to mechanical deformations is
currently a subject of high
interest~\cite{CPC,Bandgapeng,stiffening,squashed,Superfine1,Tzahi,SWNTSuperfine}.
Torsion-induced conductance oscillations have been recently
reported in multi-wall nanotubes (MWNTs), but their interpretation
left several questions unanswered~\cite{Tzahi}. To use CNTs as NEMS
elements like torsional springs and gyroscopes, it is important to
understand the origin of conductance changes induced by twisting
the nanotube.

\begin{figure}
\includegraphics[width=0.85\columnwidth]{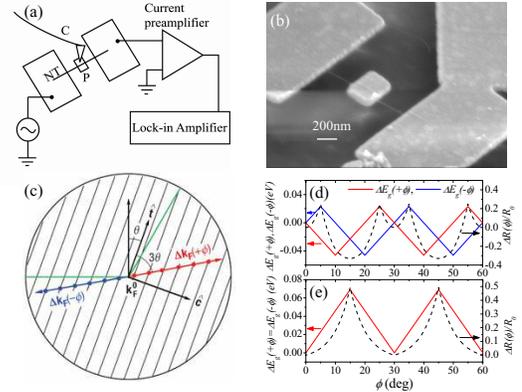}
\caption{\label{fig:figure1}(Color online) (a) Schematic of the
set-up. C is the AFM cantilever, NT the nanotube, P the pedal. (b)
SEM image of the carbon nanotube-pedal device. (c) Schematic of
the shifting of the Fermi momentum ${\bf k_F}$ with torsion. The
boundary near ${\bf k_F}$, located at a corner of the first
Brillouin zone of graphene, is indicated by the green lines. Black
parallel lines are the allowed {\bf k} sub-bands of the nanotube.
The red and blue lines show the shift in ${\bf k_F}$ for the left
and right segments of the nanotube, respectively. (d), (e) Left
axis: Change in the band gap with torsion (solid line) for the two
halves of the nanotube. Right axis: The relative resistance change
(dashed line) due to the band gap change for (d) a semiconducting
and (e) a metallic nanotube. }
\end{figure}

Here we combine electromechanical measurements with {\em ab
initio} density functional calculations to settle the controversy
about the origin of torsion-induced conductance oscillations in
multi-wall carbon nanotubes. The first explanation, referring to a
single-wall nanotube, suggests that conductance oscillations occur
due to changes in the fundamental band gap, as the Fermi momentum
${\bf k_F}$ crosses {\bf k} sub-band quantization lines, shown in
Fig.~\ref{fig:figure1}(c), while the nanotube is
twisted~\cite{YangHan2000,DT140,Tzahi}. The application to MWNTs
is justified by our finding that differential torsion effectively
decouples the walls on a MWNT near the Fermi level. The second
explanation invokes the change in registry, as the outermost wall
of a MWNT is rotated~\cite{KwonPRB} or
twisted~\cite{MWNTgraphite,RMPreview} with respect to the interior
walls. We find the first explanation to be correct, as it predicts
the conductance oscillation period to change with the inverse
square diameter of the nanotube, in accordance with our
measurements, rather than the inverse diameter, as suggested by
the latter model.

A schematic of the experiment is shown in
Fig.~\ref{fig:figure1}(a). Figure~\ref{fig:figure1}(b) shows a
fabricated suspended MWNT-pedal device with contacts at both
nanotube ends~\cite{Tzahi}. The pedal is pressed by an AFM tip,
seen in the schematic, to twist the suspended nanotube. The torque
and torsional strain on the nanotube and its conductance are
measured simultaneously as the nanotube is twisted. Here we
present results for nanotubes of different diameter to answer open
questions about the torsional electromechanical response.

\begin{figure}
\includegraphics[width=0.8\columnwidth]{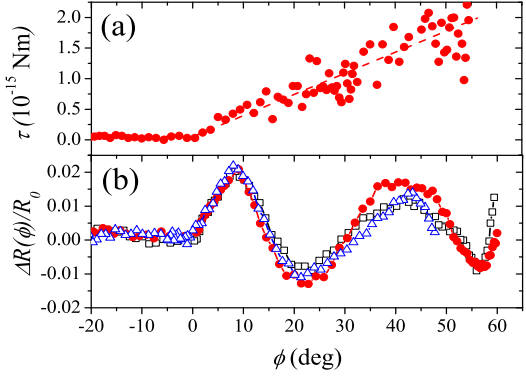}
\caption{\label{fig:figure2}(Color online) Reproducible
conductance oscillations in a nanotube of diameter 18~nm induced
by torsion. (a) Torque and (b) relative resistance change as a
function of the torsion angle $\phi$.}
\end{figure}

The torque $\tau$ experienced by the nanotube as a function of the
torsion angle $\phi$ is plotted in Fig.~\ref{fig:figure2}(a).
Figure~\ref{fig:figure2}(b) shows the oscillatory behavior in
the relative resistance change $\Delta R(\phi)/R_0$ as the
nanotube gets twisted, with $R_0=R(\phi=0)$ as the reference. The
observed behavior is reproducible for several press-retract
cycles.


\begin{figure}[b]
\includegraphics[width=0.85\columnwidth]{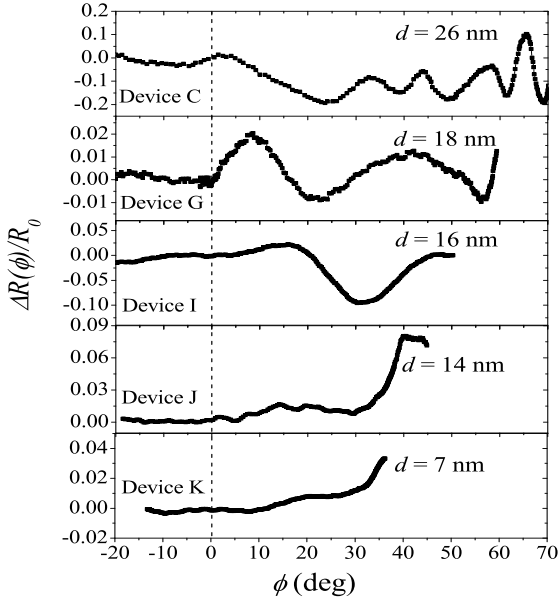}
\caption{\label{fig:figure3} Relative resistance change as a
function of the torsion angle for nanotubes of different diameter
$d$.}
\end{figure}

\begin{figure}[t]
\includegraphics[width=0.8\columnwidth]{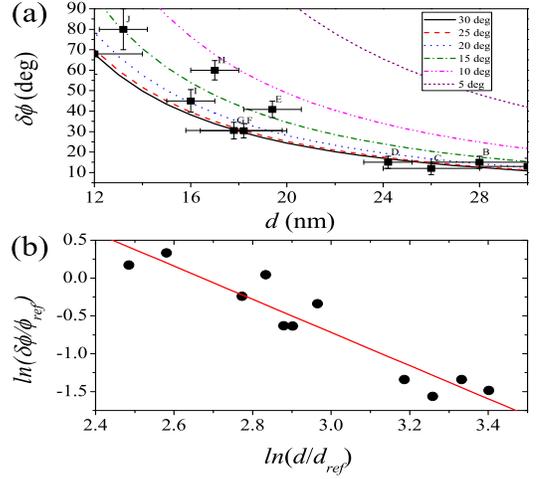}
\caption{\label{fig:figure4}(Color online) (a) Dependence of the
oscillation period $\delta\phi$ on the diameter $d$. Experimental
data $\delta\phi_{obs}$ are shown by squares, theoretical
predictions for different chiral angles are shown by lines. (b)
Double-logarithmic plot of $\delta\phi/\phi_{ref}$ as a function of
$d/d_{ref}$, with $\phi_{ref}=1$~rad and $d_{ref}=1$~nm. Our data can be
fitted by a straight line with a slope of $-2.19{\pm}0.25$.}
\end{figure}

$\Delta R(\phi)/R_0$ as a function of torsion angle is shown in
Fig.~\ref{fig:figure3} for a few representative MWNT diameters.
The gradual increase in the oscillation period $\delta\phi$ with
decreasing nanotube diameter is evident. The oscillation period
for every MWNT was obtained by taking an average from several
press-retract cycles. The observed oscillation period
$\delta\phi_{obs}$ is represented by solid squares in
Fig.~\ref{fig:figure4}(a) as a function of the MWNT diameter $d$.

We first examine the consequences of the shift in ${\bf k_F}$ on
the nanotube conductance. For a nanotube under torsion, the shift
in ${\bf k_F}$ relative to the invariant {\bf k} lines in the
circumferential direction is given by
${\Delta}k_F^c={\phi}d\sin(3\theta_0)/
(2ld_{C-C})$~\cite{YangHan2000,Tzahi}.
Here $l$ is the length of the twisted section of the nanotube and
$d_{C-C}$ is the carbon-carbon bond length. $\theta_0$ is the
chiral angle or chirality of the untwisted nanotube, defined as
the angle between the chiral vector and the zigzag direction on a
graphene plane.

Due to the linearity of the dispersion relation near ${\bf k_F}$,
the band gap $E_g$ is proportional to the distance of ${\bf k_F}$
in the circumferential direction from the nearest {\bf k}
sub-band. Therefore, $\Delta E_{g}$ initially changes in a linear fashion
with torsion. However, when ${\bf k_F}$ reaches one of the ${\bf k}$ lines,
the band gap vanishes and further torsion causes renewed opening
of the gap. The band gap reaches its maximum at the midpoint
between two ${\bf k}$ sub-bands, where further torsion decreases
its value. Thus, the band gap oscillates between zero and its
maximum value, leading to periodic metal-semiconductor transitions
(M-S effect). The change in band gap with torsion is thus given by
\begin{equation}
\label{bandgapeq} \Delta E _{g} = \frac{6\gamma_0d_{C-C}}{d}
\biggl[ \frac{d^2\sin(3\theta_0)}{4d_{C-C}l}\phi+\Delta j \biggr]
\;,
\end{equation}
where $\Delta j$ is an integer corresponding to the change in the
quantum number associated with the nearest sub-band before and
after twisting. Simulated $\Delta E_{g}$ for a CNT of $d=18$~nm is
plotted in Figs.~\ref{fig:figure1}(d) and \ref{fig:figure1}(e).
Figure~\ref{fig:figure1}(d) shows $\Delta E_{g}$ for an initially
semiconducting nanotube. Since the two halves of the CNT
experience opposite torsion angles when the pedal is pressed,
${\Delta}E_{g}$ for each half is shown separately.
Figure~\ref{fig:figure1}(e) shows $\Delta E_{g}$ for an initially
metallic CNT, in which case the change in the band gap is the same
for both halves.

The $\Delta R(\phi)/R_0$ resulting from the $\Delta E_{g}$ is
given by
\begin{eqnarray}
\frac{\Delta R(\phi)}{R_0} & = &
\frac{1}{A}\biggl\{\frac{1}{2}\Bigl[e^{\Delta E(\phi)/k_BT +
\Delta E(-\phi)/k_BT}\Bigr]-1\biggr\}\nonumber \;,\\
A & \equiv & 1 + \biggl(\frac{4e^2|t|^2R_c}{h} + 1
\biggr)e^{-E_0/k_BT} \;,
\end{eqnarray}
where $\Delta E(\phi)$ is the torsion-induced change in the
activation energy, equal to $\Delta E_g/2$. A is an attenuation
factor, $R_c$ is the contact resistance, $h$ is Planck's constant
and and $e$ the electron charge. $|t|^2$ is the transmission
probability~\cite{Tzahi} and $E_0$ is the initial activation
energy. Thus, $\Delta R(\phi)/R_0$ has the same oscillation period
as $\Delta E_g$. This can also be seen from
Fig.~\ref{fig:figure1}(d) and \ref{fig:figure1}(e), where
${\Delta}R(\phi)/R_0$ is shown by the dashed lines.

For a nanotube of diameter $d$, the sub-band spacing is
${\delta}k=2/d$. Therefore, the conductance oscillations occur
with the period ${\Delta}k_F^c=2/d$, which translates to
\begin{equation}
\label{deltaphi} \delta\phi =
\frac{4}{d^2}\frac{ld_{C-C}}{\sin(3\theta_0)} \;.
\end{equation}

This equation provides the following insight:

(i) The minimum oscillation period for a nanotube of a particular
diameter is given by $\delta\phi_{min}=4ld_{C-C}/d^2$, which is
shown by the solid line in Fig.~\ref{fig:figure4}(a). The figure
clearly indicates that $\delta\phi_{obs}>\delta\phi_{min}$ within
experimental error.

(ii) When a zigzag nanotube with $\theta_0=0$ is twisted, ${\bf
k_F}$ moves parallel to the ${\bf k}$ lines, thus keeping the band
gap constant. This also follows from Eq.~(\ref{deltaphi}), which
suggests that $\delta\phi\rightarrow\infty$ as
$\theta_0\rightarrow{0}$. However, assuming a homogeneous
distribution of chiralities, we can expect ${\approx}83$\% of the
nanotubes to have chiralities in the range 5$^\circ$ to 30$^\circ$
and $\approx$ 67$\%$ to have chiralities between 10$^\circ$ and
30$^\circ$. Also plotted in Fig.~\ref{fig:figure4}(a) by various
broken lines is the theoretical value of $\delta\phi$ according to
Eq.~(\ref{deltaphi}) for chiralities between 5$^\circ$ and
30$^\circ$. All experimental points lie in the chirality range
10$^\circ$ to 30$^\circ$. We speculate that this could simply be
because the chirality distribution may not be homogeneous and
higher chiralities are favored during the MWNT growth. Another
possible reason is that strain-induced displacements of the
triangular sublattices in graphene, known to change the band
gap~\cite{bicont}, have been neglected.
We thus expect even zig-zag nanotubes to change their conductance
with torsion. This could mean that our estimated nanotube
chiralities may be higher than the actual values.


(iii) According to Eq.~(\ref{deltaphi}), a plot of
$\ln(\delta\phi/\phi_{ref})$ versus $\ln(d/d_{ref})$ should give a
straight line with the slope $-2$. A linear fit of these data,
shown in Fig.~\ref{fig:figure4}(b), indicates an optimum slope of
$-2.19{\pm}0.25$, supporting our claim that the oscillation period
$\delta\phi$ is proportional to $1/d^2$.

An alternative explanation of the conductance oscillations links
them to changes in registry between the nanotube walls of a MWNT
as the outermost wall is twisted. This could modify the electronic
coupling between the walls, causing changes in conductance.
Considering only the outermost and the neighboring inner wall, the
double-wall nanotube structure forms a Moir\'{e} pattern, which
can be thought of as beats in two dimensions. Therefore, for a
particular shear strain $\xi$, the number of coincidences in the
Moir\'{e} pattern varies linearly with $\xi$. $\delta\phi$ should
then be inversely proportional to the number of these
coincidences, $\delta\phi\propto{1}/\xi$. Since $\xi\propto{d}$,
we would expect $\delta\phi\propto{1}/d$. Our observation of a
different functional dependence, $\delta\phi\propto{1}/d^2$, thus
excludes the Moir\'{e} effect as a cause of the conductance
oscillations.

In order to understand the apparent absence of the Moir\'{e}
effect and justify neglecting the inter-wall interactions when
determining the band gap, we performed geometry optimization and
electronic structure calculations of corresponding model systems.
We used the {\em ab initio} density functional theory (DFT)
formalism in the local density approximation (LDA), as implemented
in the SIESTA code \cite{SolerJPhysCM2002}, with a double-$\zeta$
basis set including polarization orbitals, and a mesh cutoff
energy of 200~Ry. We used an ultra-fine $k$-point mesh equivalent
to a $201{\times}201$ $k$-point sampling of the graphene Brillouin
zone, including the $\Gamma$ point. All structures were optimized
until all force components on atoms were less than 0.01~eV/{\AA}.
The calculated in-layer bond length $d_{C-C}=1.42$~{\AA} and the
interlayer spacing $c=3.34$~{\AA} in the optimized structure of
bulk hexagonal graphite showed a deviation of no more than $1$\%
from the experimental data \cite{BTKelly1981}.

\begin{figure}
\includegraphics[width=0.8\columnwidth]{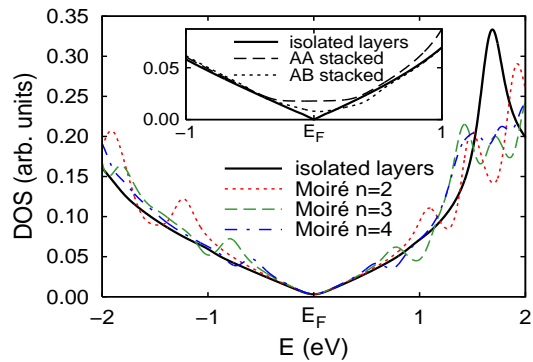}
\caption{\label{fig:figure5} (Color online) Electronic density of
states (DOS) of graphene bilayers. Near the Fermi level, the DOS
of isolated layers (solid line) agrees well with that of bilayers
forming Moir\'{e} patterns with $n=2,3,4$. The DOS of AA and AB
stacked bilayers is compared to that of isolated layers in the
inset. }
\end{figure}

Before addressing twist-related changes in the transport of MWNTs,
we investigate the effect of inter-wall interaction on the
electronic structure of double-wall nanotubes (DWNTs), which,
especially in the limiting case of very large diameters, should
closely resemble that of graphene bi-layers. Since it is extremely
unlikely to find $(n_1,m_1)@(n_2,m_2)$ DWNTs with chiral indices
representing a commensurate structure of the adjacent walls, as it
can be found in AA or AB stacked graphene bilayers, we will focus
on the general case of incommensurate DWNT structures. The
counterpart of a DWNT with incommensurate $(n_1,m_1)$ and
$(n_2,m_2)$ walls is a graphene bilayer, where the upper layer has
been rigidly rotated--within its plane--with respect to the layer
below. Both in DWNTs and graphene bilayers, for specific rotation
angles or chiral index combinations, we commonly find
quasi-commensurate arrangements with large but finite unit cells,
associated with Moir\'{e} patterns. With the graphene layers in a
bilayer being spanned by the primitive lattice vectors ${\bf a_1}$
and ${\bf a_2}$, a Moir\'{e} pattern can be produced by rotating
the initially coinciding layers about one site so that the lattice
point $n{\bf a_1}+(n-1){\bf a_2}$ in the upper layer aligns with
the lattice point $(n-1){\bf a_1}+n{\bf a_2}$ in the lower layer.

In Fig.~\ref{fig:figure5} we compare the density of states (DOS)
of a graphene monolayer to that of AA and AB stacked bilayers and
bilayers forming Moir\'{e} patterns with $n=2,3,4$. Our results
show that a graphene monolayer is a zero-gap semiconductor,
whereas commensurate AA or AB bilayers are metallic. The cause is
the finite inter-layer coupling, combined with the favorable
symmetry in the latter systems, which introduces new states near
$E_F$. Due to the lack of such symmetry in graphene bilayers
forming Moir\'{e} patterns, we find no signature of the
inter-layer coupling in the electronic structure near the Fermi
level. As seen in Fig.~\ref{fig:figure5}, the electronic density
of states of such incommensurate structures is nearly identical to
that of isolated graphene monolayers. This finding holds even in
view of the inter-layer interaction and distance, which undergo a
modulation of up to $3$\% due to the presence of Moir\'{e}
patterns. Using our graphene-nanotube analogy, we conclude that
the electronic structure of individual walls in a MWNT is
decoupled from that of adjacent walls in the most common case of
incommensurability.

Consequently, changes in transport properties of multi-wall
nanotubes, contacted only at the outermost wall in the present
study, can be understood by ignoring the presence of interior
walls, except for their structural support that prevents
deformation or collapse of the outermost wall. In particular, the
observed conductance changes in twisted MWNTs can be interpreted
using the formalism of Yang and Han, developed for an isolated
SWNT with the diameter of the MWNT \cite{YangHan2000}.

In summary, we identified the diameter dependence of the
conductance oscillations observed in multi-wall nanotubes subject
to torsion, which allowed us to settle the controversy about the
origin of this effect. Our observations indicate that the
oscillation period is proportional to $1/d^2$, as expected for a
single-wall nanotube of the same diameter. This is supported by
our theoretical finding that differential torsion effectively
decouples the walls of a multi-wall nanotube near the Fermi level
while moving the Fermi momentum across quantization lines, thus
periodically opening and closing the fundamental gap. We could
exclude the alternative explanation of the conductance
oscillations linked to registry changes between the walls, since
it would cause a $1/d$ dependence of the oscillation period in
deviation from our data.

We thank A. Yoffe and S. R. Cohen for assistance with the clean
room and AFM respectively. This research was supported by the
Israel Science Foundation, the Kimmel Center for Nanoscale
Science, the Israeli Ministry of Defense, Minerva Stiftung, and
the Djanogly and Alhadeff and Perlman foundations. EJ holds the
Victor Erlich Career Development Chair. KSN acknowledges Feinberg
Graduate School for support. DT and SB were supported by the
National Science Foundation under NSF-NSEC grant 425826 and
NSF-NIRT grant ECS-0506309.

\end{document}